\documentclass[12pt]{iopart}
\usepackage{epsfig}  
\begin{document}

\title{What statistics can tell us about strategy in tennis}

\author{I. Y. Kawashima}
\address{Escola Paulista de Medicina - UNIFESP, 04023-062, S\~ao Paulo, SP, Brazil}
\author{O. Helene}
\address{Instituto de F\'\i sica da Universidade de S\~ao Paulo, C.P. 66318, CEP 05315-970, 
S\~ao Paulo, Brazil}
\author{M. T. Yamashita}
\address{Instituto de F\'\i sica Te\'orica, UNESP - Univ Estadual Paulista, C.P. 70532-2, 
CEP 01156-970, S\~ao Paulo, SP, Brazil}
\author{R. S. Marques de Carvalho}
\address{Departamento de Inform\'atica em Sa\'ude - Escola Paulista de Medicina - UNIFESP, 
04023-062, S\~ao Paulo, SP, Brazil}
\ead{marques.carvalho@unifesp.br}

\begin{abstract}
In this paper we analyse tiebreak results from some tennis players in order to investigate 
whether we are able to identify a non-aleatory distribution of the points in this crucial 
moment of the game. We compared the observed results with a binomial distribution considering 
that the probabilities of winning or losing a point are equal. Using a $\chi^2$ test we found 
that, excepting some players, the greatest part of the results agrees with our hypothesis 
that the points in tiebreaks are merely aleatory.
 
\end{abstract}

\vspace{2pc}

\maketitle

\section{Introduction}

A recurrent question in a signal analysis is whether it is a true signal or just a 
noise \cite{otaviano,razak}. This question arises, for example, when we are analysing 
a tomography or X-ray picture \cite{mylott} or searching for a new particle like Higgs 
boson \cite{higgs}. In these cases different statistical tests are made and usually 
the discussion is how many standard deviations we can accept or reject a given hypothesis. 
Statistical analysis of experimental data are made since the first years of physics and 
engineering courses \cite{peterlin}. The connection of classroom problems with daily 
problems \cite{esportes,tsunami} may be more stimulating than, for example, roll many 
dice hundreds times to see in practice a binomial distribution. The aim of this paper 
is to investigate whether the points in a tiebreak originate from a statistical 
fluctuation and are randomly decided.

A tennis match is divided in sets and games. To win a set the player should complete six games 
with at least two games of difference from the other player (6x0, 6x1, $\dots$, 6x4). In the 
case of a player with six games and the other with five, it is played one more game and then it may 
happen two things: if the player with six games wins the game then the set ends in 7x5. If the 
player loses the game then the set is tied and they will play a tiebreak \footnote{note that we are not 
considering the last set of Grand Slam events or Davis Cup where the games can continue 
infinitely}. During the tiebreak, the player who wins the first seven points with at least two 
points of difference of the other player wins the game and the set. If necessary, the tiebreak
continues until the minimum difference of two points is achieved.

In this paper we collected results from tiebreaks of several players. We then plotted in a 
histogram the difference of points, where positive values mean victories and negative 
losses. These histograms are compared with a theoretical binomial distribution, constrained to 
the tennis rules described in the last paragraph, but considering equal probabilities of winning 
or losing a point. We performed a $\chi^2$ test to have an objective parameter to say if 
the observed and calculated results are statistically different.

The paper is organized as follows. In section \ref{methods} we explain the 
criterium we used to select the players and the $\chi^2$ test. In section \ref{results} 
we compare the observed results and our theoretical prediction. Finally, in section \ref{conclusions} 
we summarize and give our conclusions.

\section{Methods}
\label{methods}

Analysing the conditions which may lead to a tiebreak we could consider two main 
reasons. The first one: both players may have a very similar game. 
Then, considering that the serve can really be considered as an 
advantage, we will have in this condition a very favorable condition for a tiebreak (every 
game of serve the player who is serving wins the game). The second reason usually occurs 
when one of the players has an amazing serve. Normally, this 
condition comes essentially from a very big height (exceptions to this fact may be found). 
Then, thanks to the height the agility of the player is seriously compromised, which makes 
that in one hand the tall player has a low probability to win the game when his opponent is 
serving, but on the other hand the opponent rarely can obtain a good return of the big 
serve. A good example for this second reason is Ivo Karlovic (211 cm) from Croatia. 
However, despite of what reason caused the tiebreak, the fact is: when the match goes to a 
tiebreak, in that moment the match was very balanced. Thus, it is not strange to think 
that each point could go randomly to any of the players.

In order to investigate tiebreak results, we selected the top ten players according to 
ATP (Professional Tennis Association) website in the last week of October, 2015. For the 
analysis we also included the player Ivo Karlovic, as his games usually go to a tiebreak. 
The tiebreak results were mainly extract from gambling sites, where we may find detailed 
results.

\begin{figure}[htb!]
\centerline{\epsfig{figure=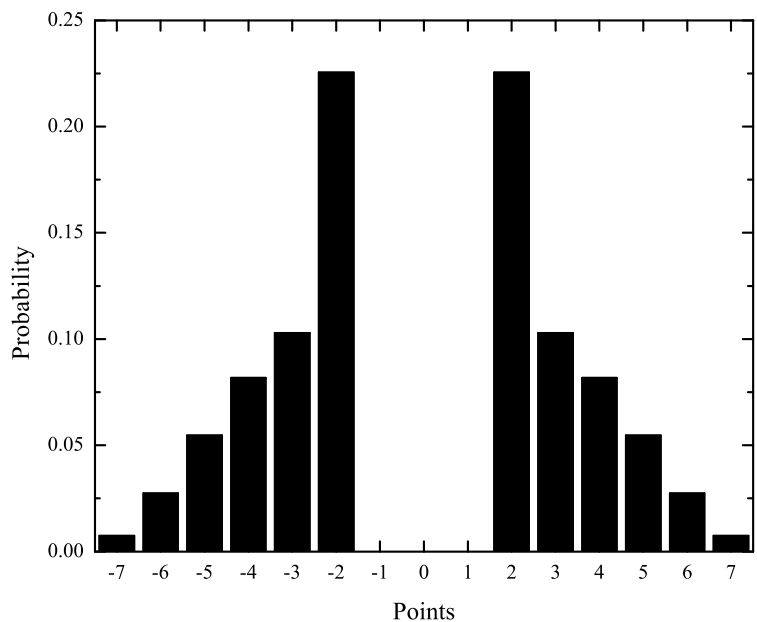,width=10cm}}
\caption[dummy0]{Theoretical result for the probability to have a result from -7 to 7 in a tiebreak 
considering that the probability to win(lose) a point is 0.5. Horizontal axis 
is the difference of points from a given and other players. Positive values mean victories and 
negative losses.}
\label{figsim}
\end{figure}

In \ref{figsim} we plotted a binomial distribution of a tiebreak in tennis 
considering that the probability to win, or lose, a point in the tiebreak is 0.5. 

We calculated the $\chi^2$ quantity in order to compare the expected results with the 
observed ones. The $\chi^2$ variable is defined as:
\begin{equation}
\label{chi}
\chi^2=\sum_{-7}^{7}\frac{(y_i^{(O)}-y_i^{(E)})^2}{y_i^{(E)}},
\end{equation}
where $y_i^{(O)}$ and $y_i^{(E)}$ are, respectively, the observed and expected number 
of events. The sum should be performed over all values. The expected number of events 
is simply given by $Np_i$, where $N$ is the total number of events and $p_i$ ($i=-7,\dots,7$) 
is the probability for result $i$ to occur (see Fig. \ref{figsim}). Defining $F=F(\chi^2)$ 
as the probability density function for eleven degrees of freedom, we may write, respectively, 
the probability of finding a smaller and a greater value than a given $\chi^2$ as
\begin{equation}
P_<=\int_0^{\chi_<^2} F(\chi^2) d\chi^2, \;\;\;\;\; 
P_>=\int_{\chi_>^2}^\infty F(\chi^2) d\chi^2.
\end{equation}

Here, we will consider that the observed values agree with our theoretical result if the calculated 
$\chi^2$ stays inside the interval $\chi_<^2<\chi^2<\chi_>^2\equiv4.575<\chi^2<19.675$, which 
corresponds to $P_<=0.05$ and $P_>=0.95$.

\section{Results}
\label{results}

\begin{figure}[htb!]
\centerline{\epsfig{figure=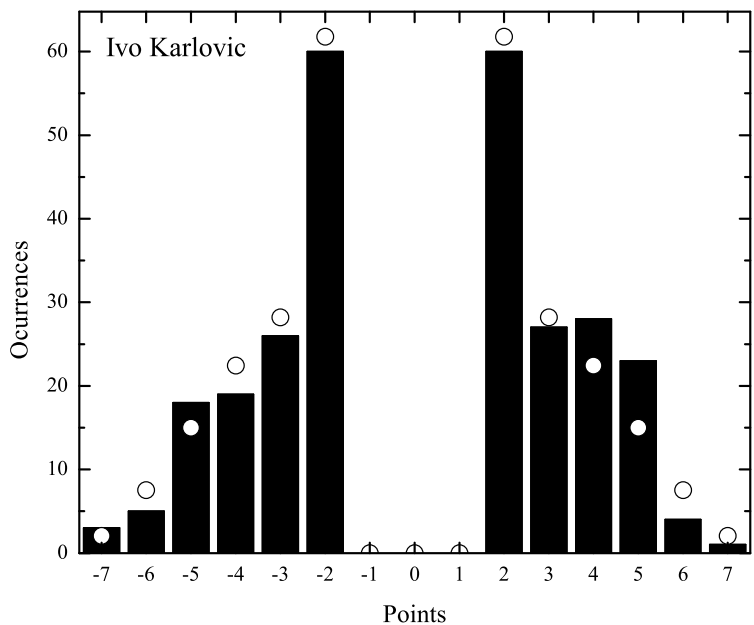,width=10cm}}
\caption[dummy0]{Histogram of Ivo Karlovic results. The bars are a total of 274 collected data 
and the open circles are our theoretical results. The agreement between both results are 
very good giving a $\chi^2=10.6$. This result means that despite the great serve from 
Karlovic, the probability of winning (or losing) a point is close to 50 \%.}
\label{karlovic}
\end{figure}

In this section we will compare our theoretical result with some observed data. Fig. \ref{karlovic}
shows the expected results, given by $Np_i$ (for Karlovic $N=274$) and represented 
by open circles, compared to the observed ones. Not only the structure of the results 
are very similar, but also the values in each channel. The calculated $\chi^2$ is 10.6, which 
is inside the interval mentioned in the last section indicating that both results are 
statistically equivalent. This means that despite the big serve from Karlovic his results 
in tiebreak are close to a completely random situation. 

\begin{figure}[htb!]
\centerline{\epsfig{figure=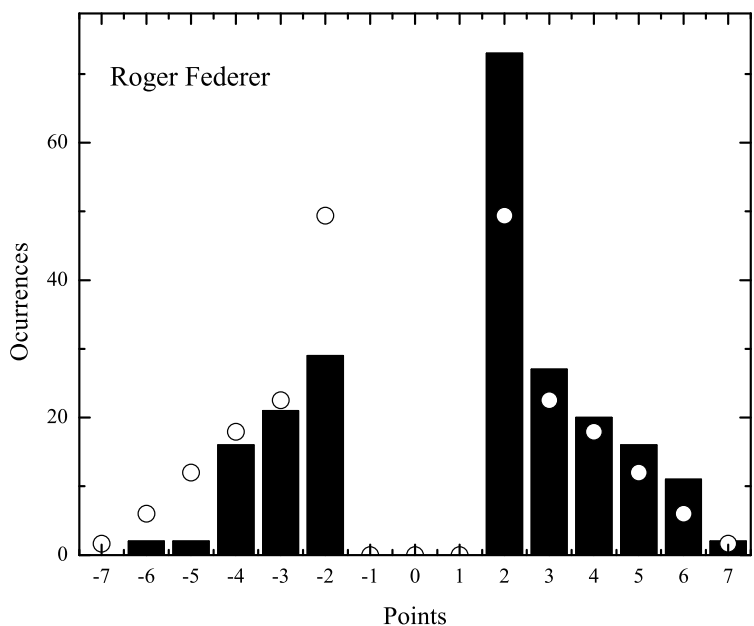,width=10cm}}
\caption[dummy0]{Histogram of Roger Federer results. Same as figure \ref{karlovic} for 
219 data and a $\chi^2=39.3$. This is a typical figure of non-aleatory results.}
\label{federer}
\end{figure}

Figure \ref{federer} shows the results from Roger Federer. We can immediately note 
the large difference from the theoretical and observed results. This is clearly a 
non-aleatory result with a $\chi^2$ exceeding by far the upper limit of 19.675. Roger Federer 
is one of the greatest tennis players of all time and this figure may demonstrate it. Tennis 
is a very mental game with moments of extreme pressure (tiebreaks, for example). 
A crucial moment occurs when the point can define the game, the set, or the match. 
The top tennis players have the capacity to increase considerably their concentration and tennis 
level in these moments. The courage to hit a drop shot or a ball down the line in a 
delicate moment avoiding the opponent to win the point is a quality that is not shared 
by all players. 

Table \ref{tablechi} shows the calculated $\chi^2$ for the top ten players at the moment 
we were writing this paper (note that the ranking changes every week) and Ivo Karlovic, who 
has the biggest number of aces in history and a large number of tiebreaks. In order to agree 
with the theoretical prediction (50 \% of probability to win or lose a point in tiebrak) 
the $\chi^2$ should stay inside the interval $[4.575,19.675]$. As we can see, Djokovic, Murray, 
Wawrinka, Ferrer, Tsonga and Karlovic agree with an aleatory result. Note that almost more 
than half of the top ranked players have practically random tiebreak results. If we 
consider lower rankings the number of players who agree with our hypothesis increases 
considerably.

\begin{table}[htb!]
\centering
\label{tablechi}
\begin{tabular}{|l|c|c|}
\hline
Player & $\chi^2$ & Data  \\
\hline
Novak Djokovic & 19.0 & 181 \\
Roger Federer & 39.3 & 219 \\
Andy Murray & 14.6 & 168 \\
Stan Wawrinka & 11.4 & 197 \\
Tomas Berdych & 21.6 & 199 \\
Rafael Nadal & 29.3 & 163 \\
Kei Nishikori & 23.7 & 115 \\
David Ferrer & 9.1 & 151 \\
Jo-Wilfried Tsonga & 11.0 & 220 \\
Milos Raonic & 31.9 & 260 \\
Ivo Karlovic & 10.6 & 274 \\
\hline
\end{tabular}
\caption{Calculated $\chi^2$-values for several players. The last column is the number 
of collected data. In order to be statistically equivalent to a random result, $\chi^2$ should 
stay inside $4.575<\chi^2<19.675$. In this table, the values larger than 19.675 correspond to 
more victories than that predicted by our model.}
\end{table}

\section{Conclusion}
\label{conclusions}

We could see from our calculations that half of the top ten players and the player who has 
the greatest number of aces in history (Karlovic) display a tiebreak result that is in 
agreement with our hypothesis of aleatory points. Definitely, this is not a statistical 
accident. The agreement with our prediction just tell that the strategy used by these players 
to play tiebreaks is returning the same result as the coin thrown in the beginning of the 
match to decide which player serves first. Besides not an easy task, the coach of these 
players could at least adopt a strategy which could arrive in a result different of 50 \%. 
Considering lower rankings the number of players who agree with our hypothesis increases 
considerably.

As written in the introduction, statistical analysis is a topic explored in the first years 
of undergraduate physics or engineering courses. A contact with a real problem where it 
is possible to analyse the data of your favorite team or player is by far more exciting 
than spend an hour (or more) rolling many dice to see in practice a binomial distribution. 
Variations of the problem treated here may be easily extended to other sports like, e.g. football. 

\ack
The authors thank PET (Programa de Educa\c c\~ao Tutorial - MEC) for support. MTY, a very good 
amateur tennis player, thanks FAPESP (Funda\c c\~ao de Amparo \`a Pesquisa do 
Estado de S\~ao Paulo) and CNPq (Conselho Nacional de Desenvolvimento Cient\'\i fico e 
Tecnol\'ogico) for partial support.

\end{document}